\documentclass[12pt]{article}
\usepackage{epsfig}

\newcommand{\beq}{\begin{equation}}
\newcommand{\eeq}{\end{equation}}
\newcommand{\beqa}{\begin{eqnarray}}
\newcommand{\eeqa}{\end{eqnarray}}
\newcommand{\beqar}{\begin{eqnarray*}}
\newcommand{\eeqar}{\end{eqnarray*}}

\newcommand{\eg}{{\it e.g.,}\ }
\newcommand{\ie}{{\it i.e.,}\ }
\newcommand{\labell}[1]{\label{#1}} 
\newcommand{\reef}[1]{(\ref{#1})}
\newcommand\prt{\partial}

\newcommand\cL{{\cal L}}

\newcommand\cZ{{\cal Z}}
\newcommand\cV{{\cal V}}

\newcommand\cE{{\cal E}}

\newcommand\Tr{{\rm Tr}}

\begin{document}

 \vspace*{1cm}

\begin{center}
{\bf \Large Tachyon Tunnelling in D-brane-anti-D-brane

 }
\vspace*{1cm}

{ Kazem Bitaghsir and Mohammad R. Garousi}\\
\vspace*{0.2cm}
{ Department of Physics, Ferdowsi university, P.O. Box 1436, Mashhad, Iran}\\
\vspace*{0.1cm}
{ Institute for Studies in Theoretical Physics and Mathematics
IPM} \\
{P.O. Box 19395-5531, Tehran, Iran}\\
\vspace*{0.4cm}

\vspace{2cm}
ABSTRACT
\end{center}
Using the  tachyon DBI action  proposal  for the effective theory
of   non-coincident D$_p$-brane-anti-D$_p$-brane system, we study
the decay of this system in the tachyon channel. We assume that
the branes separation is held fixed, \ie no throat formation,  and
then find the bounce solution which describe the decay of the
system from false to  the true vacuum of the tachyon potential.
We shall show that due to the non-standard form of the kinetic
term in the effective action, the thin wall approximation for
calculating the bubble nucleation rate gives a result which is
independent of the branes separation. This unusual result might
indicate that the true decay of this metastable system  should be
via a solution that represents a throat formation as well as the
tachyon tunneling.


\vfill \setcounter{page}{0} \setcounter{footnote}{0}
\newpage

\section{Introduction} \label{intro}
\begin{figure}
  \begin{center}
 \includegraphics{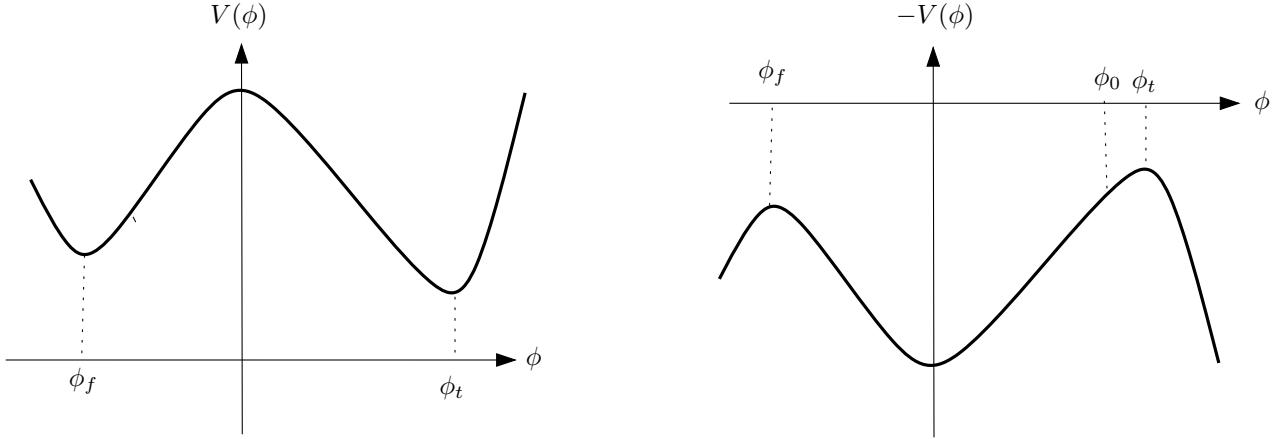}
    \end{center}
 \begin{quote}
  \caption{\it A potential $V(\phi)$ which has two
  non-degenerate vacuum $\phi_f$
  and $\phi_t$ with $V(\phi_t)<V(\phi_f)$, and
  potential $-V(\phi)$ which appears in the
  equation of motion of the Euclideanized action. }
  \end{quote}
  \label{Fig1}
\end{figure}

It was shown by Coleman \cite{Coleman} that in a scalar field
theory with potential $V(\phi)$ which has both false vacuum,
$\phi_f$, and true vacuum, $\phi_t$ (see fig.\ref{Fig1}),  and
with standard kinetic term, \ie \beqa S&=&-\int d^{p+1} x
\left(\frac{1}{2}(\prt_a\phi)^2+V(\phi)\right)\,,\label{s1}\eeqa
where  $a$ is a world volume index, there is always a vacuum
tunneling from the false vacuum to the true vacuum. The false
vacuum decays via a quantum mechanical tunneling process that
leads to the nucleation of bubbles of true vacuum. The
semiclassical calculation of the bubble nucleation rate per unit
volume, $\Gamma$, is given by \cite{Coleman}\beqa \Gamma
&\sim&e^{-B}\labell{drate}\,,\eeqa where B is obtained from the
action of the bounce solution to the Euclideanized field
equations. The "overshoot" argument of Coleman \cite{Coleman}
guarantees the existence of the bounce solution. This argument is
the following: Consider the equation of motion of the
Euclideanized action  for maximal symmetric solution and for
$p>0$ \beqa \ddot{\phi}+\frac{p}{r}\dot{\phi}=V'\,.\eeqa This
equation is like the equation of motion of a particle in the
potential $-V$ with the time dependent kinetic friction term
$p\dot{\phi}/r$ (see fig.\ref{Fig1}). The bounce solution is a
solution of the above equation that starts  with the initial
condition $ \dot{\phi}(0)=0$ and $\phi(0)=\phi_0$, and  approaches
$\phi(\infty)=\phi_f$ with zero velocity. Around the maximum of
$-V$ one can safely write $V'(\phi)\sim \mu^2(\phi-\phi_t)$,
where $\mu^2=V''(\phi_t)>0$, hence the above equation can be
written in the linear form \beqa
\ddot{\phi}+\frac{p}{r}\dot{\phi}&=&\mu^2 (\phi-\phi_t)\,.\eeqa
In terms of new variable $(\phi-\phi_t)=r^{(1-p)/2}\psi$, this
equation converts to Bessel equation \beqa
\ddot{\psi}+\frac{1}{r}\dot{\psi}-\left(\mu^2+
\frac{(1-p)^2}{4r^2}\right)\psi&=&0\,,\eeqa whose solution is
Bessel function $\psi(r)=I_{(p-1)/2}(\mu r)$. Hence the solution
of $\phi$ is\beqa
\phi(r)-\phi_t&=&\Gamma(\frac{p+1}{2})(\phi(0)-\phi_t)I_{(p-1)/2}(\mu
r)\left(\frac{2}{\mu r}\right)^{(p-1)/2}\,\,.\eeqa
 If initially $\phi$ is very close to
$\phi_t$, it will stay there for long time. After that time it
rolls  with negligible  friction term and with finite velocity
down the potential $-V$ toward the vacuum $\phi_f$ which is at
lower energy. Hence, there is always an "overshoot" point for any
potential with both false and true vacuum.  An approximated bounce
solution (the thin-wall approximation) can be found for any
potential \cite{Coleman}. In this approximation, the particle
stays at true vacuum for a fixed period of time. Then it moves
quickly without friction  through the valley of the potential
$-V$, and slowly comes to rest at false vacuum at time infinity.
The form of this solution depends on details of the potential $V$.

A physical system in string theory that should be  described
effectively by a field theory that has potential with both false
and true vacuums, is the non-coincident D-brane-anti-D-brane
system  \cite{Banks}. Apart from the complex tachyon  which is a
scalar field that has a potential with both false and true
vacuums, the world-volume of this metastable system has massless
transverse scalar  fields as well.  It has been pointed out in
\cite{callen} that the true decay channel for this system is the
following: the branes attract each other by long range
gravitational forces, and then they annihilate each other via a
direct appearance of tachyon instability. However, when branes
separation is much larger than the string length scale, this
system may decay in another way: through the tunnel effect by
creation of a throat between the branes \cite{callen}.  Assuming
that the tachyon is frozen at the false vacuum, the authors have
found a throat solution to the Euclideanized equation of motion
of the massless scalar fields. The decay rate for nucleating this
throat between the two branes was found to be given by \beqa
B_{scalar}&=&T_p\Delta^{p+1}\frac{\sqrt{\pi}}{\Gamma\left(\frac{p+3}{2}\right)}
\left(\frac{p\Gamma(1-1/2p)}{\Gamma(1/2-1/2p)}\right)^p\,,\labell{Bscalar}\eeqa
where $ \Delta$ is the branes separation. This result truly
indicates that as the branes separation goes to infinity the
decay rate \reef{drate} goes to zero. It has been noted in
\cite{callen} that if the branes are free to move in the space,
the time scale of the gravitational approach for the decay is much
smaller than that of the decay via the throat formation. However,
there are situations where the latter time scale is dominate
\cite{koji}.

Having a scalar field in the effective action of $D\bar{D}$ which
has both false and true vacuums, \ie the tachyon field, it is
natural to ask what happens to this system when tachyon tunnels
from the false vacuum to the true vacuum. We are then interested
in the bounce solution of the Euclieanized equation of motion of
the tachyon field. In the first step, however, we assume that the
branes are frozen at specific position in the transverse space,
\ie no throat formation,  and consider only the tachyon as
dynamical field. This bounce solution has been studied in
\cite{koji} using two-derivative truncation of the BSFT effective
action. In the present paper, however, we use the tachyon DBI
action, proposed in \cite{sen3, MRG} for describing effectively
the non-coincident brane-anti-brane system,
 to study this bounce solution. We shall find
that in the thin-wall approximation, the decay rate for
nucleating the bubble  is given by \beqa
B_{tachyon}&=&T_p(\ell_c)^{p+1}\frac{\sqrt{\pi}}
{\Gamma\left(\frac{p+3}{2}\right)}(\sqrt{\pi}p/2)^p\eeqa where
$\ell_c=\sqrt{2\pi^2\alpha'}$. Unlike the result in
\reef{Bscalar}, the above  is independent of the branes
separation! We interpret   this unusual result as a sign that the
assumption that there is no throat formation between the two
branes  is not a valid assumption. In other words, the true decay
of the metastable  $D\bar{D}$ system should be via a solution of
the coupled equation of motion of massless scalar and tachyon
which represents both tachyon bounce and scalar throat formation.

In the next section, we review the construction of the tachyon
DBI action proposed in \cite{sen3,MRG} for the effective theory of
non-coincident $D_p\bar{D}_p$ system. In section 3, using the
assumption  that the branes separation is constant, we study the
bounce solution of the tachyon equation  and calculate the bubble
nucleation rate for large brane separation.

A   tachyon DBI  field theory has also been used to study the
vacuum tunneling from false to the true vacuum \cite{LC}. The
physical process studied in \cite{LC} is nucleation of spherical
D-branes in the presence of an external RR electric field
\cite{CT}. Using the description of D-branes as solitons of the
tachyon field on non-BPS D-branes, they calculated the rate of
spherical D-branes nucleation  as the tachyon tunneling from
false to the true vacuum and find exact agreement with the result
in \cite{CT}.

\section{Review of construction of $D_p\bar{D}_p$ effective action}

In this section we review the construction of  the  effective
action of fixed D$_p$-brane-anti-D$_p$-brane proposed in
\cite{MRG}. Consider the effective action of one fixed non-BPS
D$_p$-brane in type IIA(IIB) theory in flat
background\cite{sen0,mg1,Berg,jk}:
 \beqa
S&=&-\int d^{p+1}x V(T) \sqrt{-\det(\eta_{ab}+\alpha'\prt_a
T\prt_b T)} \,\,,\labell{dbiac2}\eeqa where
$V(T)=T_p(1-T^2/4+O(T^4))$ is the tachyon potential. A potential
which is consistent with S-matrix element calculation up to $T^4$
term  is $V(T)=T_p\,e^{-T^2/4}$ \cite{RG}. This potential appears
also in the BSFT tachyon effective action \cite{PK}.

The  kink solution  of tachyon should be the BPS D$_{p-1}$-brane
of type IIA(IIB) \cite{sen5}. The tension of the kink is given by
$T_{p-1}= \sqrt{\alpha'}\int_{-T_0}^{T_0}V(T)dT$ where $T_0$ is
the value of the tachyon potential at its minimum. There are many
different tachyon potentials which correctly reproduce the
tension of the BPS brane \cite{sen3,ali}, \ie
$T_{p-1}=\pi\sqrt{2\alpha'}T_p$. One example is the following
potential \cite{kim,nlhl}: \beqa V(T)=\frac{T_p}{\cosh(T/\sqrt{2}
)}\,\,.\labell{tacpo}\eeqa This has minimum at $T\rightarrow
\pm\infty$ and behaves as $V(T)\sim e^{-T/\sqrt{2}}$ at
$T\rightarrow \infty$. This potential is also consistent with the
fact that there is no open string state at the end of the tachyon
condensation \cite{asen3}.

The  effective action of  $N$ non-BPS D-branes should be the
non-abelian extension of the effective action of one non-BPS
D-brane \cite{mg1}. For two fixed non-BPS D-branes and for trivial
background, the effective action is \cite{mg1}\beqa S&=&-\int
d^{p+1}x \Tr\left(
V(T)\sqrt{\det(Q)}\sqrt{-\det\left(\eta_{ab}+
T_{ab}\right)}\right)\,,\label{s2}\eeqa
where the matrix  $T_{ab}$ is  \beqa
T_{ab}&=&\alpha'\prt_aT\prt_bT+\frac{1}{2\pi}
\prt_aT[X^i,T](Q^{-1})_{ij}[X^j,T]
\prt_bT \,,\nonumber\eeqa and matrix $Q^i{}_j$ is \beqa
Q^i{}_j&=&I\delta^i{}_j-\frac{i}{2\pi\alpha'}[X^i,X^k]\eta_{kj}
-\frac{1}{(2\pi)^2\alpha'}[X^i,T][X^k,T]\eta_{kj}\,.\eeqa The
trace in the action \reef{s2} should be completely symmetric
between all non-abelian expressions of the form
$[X^i,X^j],\prt_aT,[X^i,T]$, and individual $T$ of the tachyon
potential.  The matrices $X^i$ and $T$ are \beqa
X^{i}=\pmatrix{X^{(11)i}&X^{(12)i}\cr
X^{(21)i}&X^{(22)i}},\,\,T=\pmatrix{T^{(11)}&T^{(12)}\cr
T^{(21)}&T^{(22)}}\,, \labell{M0}\eeqa where superscripts
$(11),(12),(21),(22)$ refer to the ends of open strings, \eg
$(12)$ means the open string with one end on brane $1$ and the
other end on brane $2$.

Now,  type IIA(IIB) converts to type IIB(IIA) under  orbifolding
by $(-1)^{F_L^S}$. In particular,  a non-BPS $D_p$-brane of
IIA(IIB) converts to BPS $D_p$-brane or anti-$D_p$-brane of
IIB(IIA) theory \cite{sen2}. Hence, two non-BPS $D_p$-branes of
IIA(IIB) may convert to $D_p$-brane-anti-$D_p$-brane of IIB(IIA).
This orbifolding converts the matrices \reef{M0} to the
following:\beqa X^{i}=\pmatrix{X^{(1)i}&0\cr
0&X^{(2)i}},\,\,T=\pmatrix{0&\tau\cr \tau^*&0}\,.
\labell{M1}\eeqa where we have changed the notation here, \ie the
superscripts $(1)$ and $ (2)$ refer to the open string fields
with both ends on brane $1$ and $2$, respectively, and  $\tau
(\tau^*)$ refers to the tachyon with one end on brane $1(2)$ and
the other end on brane $2(1)$.  The above matrices satisfy  the
following relations: \beqa &[X^i,X^j]=0,\,\,\,
&[X^i,T]=\ell^i\pmatrix{0&\tau\cr - \tau^*&0},\,\,\,\nonumber
\eeqa where $\ell^i=X^{(1)i}-X^{(2)i}$ is the field that
represents the distance between the two branes. We have assumed,
however, that this field is constant. The matrix $Q^i{}_j$
simplifies to \beqa Q^i{}_j&=&I\left(\delta^i{}_j+
\frac{|\tau|^2}{(2\pi)^2\alpha'}\ell^i\ell_j\right)\,.\nonumber\eeqa
The inverse of this matrix is \beqa
(Q^{-1})^i{}_j&=&I\left(\delta^i{}_j-\frac{|\tau|^2}{(2\pi)^2\alpha'\det(
Q) }\ell^i\ell_j\right)\,, \eeqa where \beqa \det(Q)&=&I\left(1+
\frac{|\tau|^2\ell^2}{(2\pi)^2\alpha'}\right)\,.\labell{Q1}\eeqa
The matrix $T_{ab}$  simplifies to \beqa
T_{ab}&=&\frac{\alpha'}{\det(Q)}\left(\prt_aT\prt_bT
\right)\,=\,\frac{\alpha'}{\det(Q)}\pmatrix{\prt_a\tau\prt_b\tau^*&0\cr
0&\prt_a\tau^*\prt_b\tau}\,\,.\labell{T1}\eeqa Note that this
matrix is not a real matrix, however, one expects to have a real
action after implementing the trace prescription.

\begin{figure}
  \begin{center}
 \includegraphics[height=2.5in,width=2.5in]{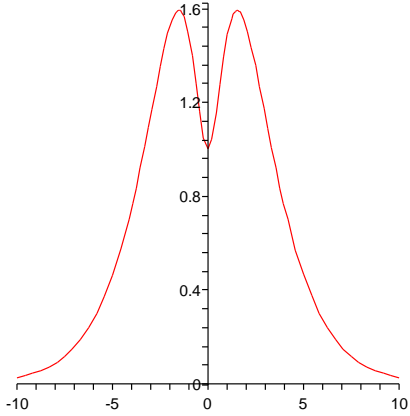}
    \end{center}

  \begin{quote}
  \caption{\it The $D_p\bar{D}_p$ potential
  $\cV(\phi,\ell)$ for brane separation
  $\ell>\ell_c$ in terms of $\phi$. }
  \end{quote}
  \label{Fig2}
\end{figure}

Inserting the expressions \reef{Q1} and \reef{T1} into \reef{s2},
and performing the symmetric trace, one finds the following
effective action for non-coincident $D_p\bar{D}_p$ system of
IIB(IIA) theory: \beqa S&=&-2\int
d^{p+1}x\,\cV(|\tau|,\ell)\left(\sqrt{-\det(\eta_{ab}+
\frac{\alpha'/2}{1+\frac{|\tau|^2\ell^2}{(2\pi)^2\alpha'}}
(\prt_a\tau\prt_b\tau^*+
\prt_a\tau^*\prt_b\tau)}\right)\label{s3}\,,\eeqa where the
tachyon  potential is \beqa
\cV(|\tau|,\ell)&=&V(|\tau|)\sqrt{1+\frac{|\tau|^2\ell^2}
{(2\pi)^2\alpha'}}\,,\eeqa where $V(|\tau|)$ is the tachyon
potential of the original non-BPS $D_p$-brane of IIA(IIB) theory,
\eg \reef{tacpo}. Note that $\cV(|\tau|,\ell=0)=V(|\tau|)$. See
\cite{Ig}, for other proposal for the effective potential of
non-coincident $D_p\bar{D}_p$ system. For $\ell>\ell_c$ where
$\ell_c\equiv\sqrt{2\pi^2\alpha'}$, the above potential has a
barrier, see fig.2. Having two non-degenerate minima for the
potential, one may expect that the $D_p\bar{D}_p$ system makes a
bubble formation by tachyon tunneling through the barrier. In the
next section we will study this tunneling.

\section{Bounce solution of $D_p\bar{D}_p$ system}
The tachyon potential depends only on the amplitude of the
complex tachyon, hence , the  phase of this field can be held
fixed. Writing the complex tachyon as $\tau=\phi e^{i\theta}$,
the action \reef{s3} for fixed $\theta$ becomes

\beqa S&=&-2\int d^{p+1}x
\cV(\phi,\ell)\sqrt{-\det\left(\eta_{ab}+\frac{\alpha'}{1+
\frac{\phi^2\ell^2}{(2\pi)^2\alpha'}}
\prt_a\phi\prt_b\phi\right)}\,\,\,,\nonumber\eeqa This action can
be rewritten as \beqa S&=&-\frac{T_{p-1}}{\sqrt{\alpha'}}\int
d^{p+1}x U(\phi)\sqrt{1+\frac{\phi^2\ell^2}{(2\pi)^2\alpha'}+
\alpha' \prt_a\phi\prt^a\phi}\nonumber\,\,\,,\eeqa where we have
rescaled  the tachyon potential by $2\sqrt{\alpha'}/T_{p-1}$ and
$T_{p-1}$ is the tension of the BPS D$_{p-1}$-brane of type
IIA(IIB), \ie $U(\phi)=2\sqrt{\alpha'}V(\phi)/T_{p-1}$. The
tachyon potential  is now dimensionless.

The Euclidean action is \beqa
S_E&=&\frac{T_{p-1}}{\sqrt{\alpha'}}\int d^{p+1}x
U(\phi)\sqrt{1+\frac{\phi^2\ell^2}{(2\pi)^2\alpha'}+ \alpha'
\prt_a\phi\prt^a\phi}\,\,\,.\label{SE}\eeqa The tachyon potential
$\cV$ has local minimum at $\phi=0$ and global minima at
$\phi=\pm\infty$. Since the true vacuums are at infinity, it is
convenient to define a new field as \beqa
\Phi&=&\int_0^{\phi}U(\phi')d\phi'\,\,.\label{newphi}\eeqa In
terms of this new field,  the false vacuum is  at $\Phi=0$ and the
true vacuums are  at $\Phi=\pm 1$. The value of the dimensionless
tachyon potential, $\cZ=2\sqrt{\alpha'}\cV/T_{p-1}$,  at false
vacuum is $\cZ(0,\ell)=\sqrt{2}/\pi$, and at the true vacuums is
$\cZ(\pm 1,\ell)=0$.

For the  potential \reef{tacpo}, the relation between $\phi$ and
$\Phi$ is $e^{\phi/\sqrt{2}}=\tan(\pi(1+\Phi)/4)$, and the
tachyon potential $\cZ$ is \beqa
\cZ(\Phi,\ell)&=&U(\Phi)\sqrt{1+\frac{\ell^2}
{(2\pi)^2\alpha'}\phi^2(\Phi)}\label{Z}\\
&=&\frac{\sqrt{2}}{\pi}\cos\left(\frac{\pi}{2}\Phi\right)\sqrt{1+
\frac{\ell^2}{2\pi^2\alpha'}\left[\ln
\left(\tan\left(\frac{\pi}{4}(1+\Phi)\right)\right)\right]^2}\nonumber\,\,.\eeqa
In fig.3, $-\cZ(\Phi,\ell)$ is plotted for different values of the
brane separation $\ell$.
 \begin{figure}
  \begin{center}
 \includegraphics[height=2.5in,width=2.5in]{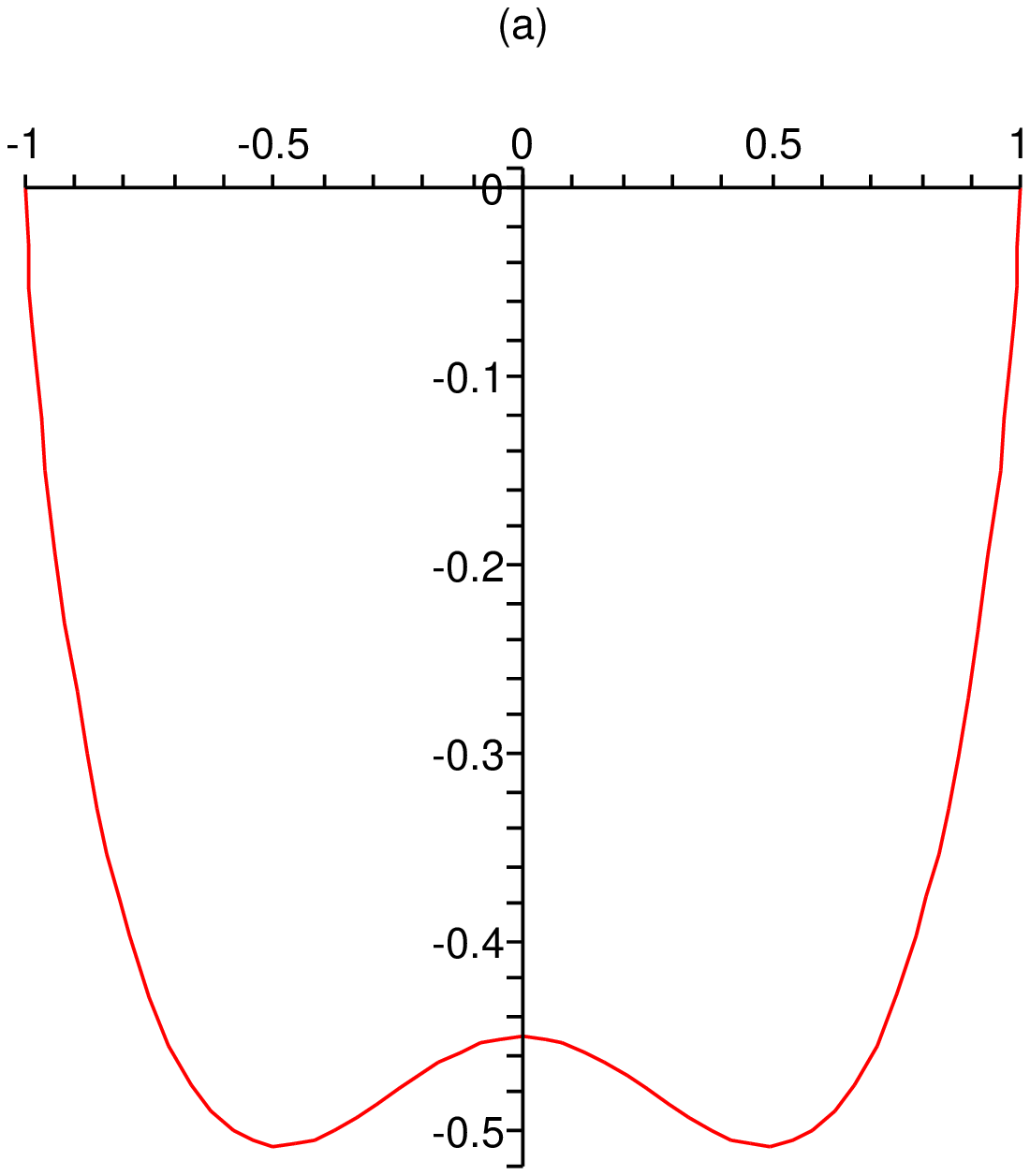}
 \includegraphics[height=2.5in,width=2.5in]{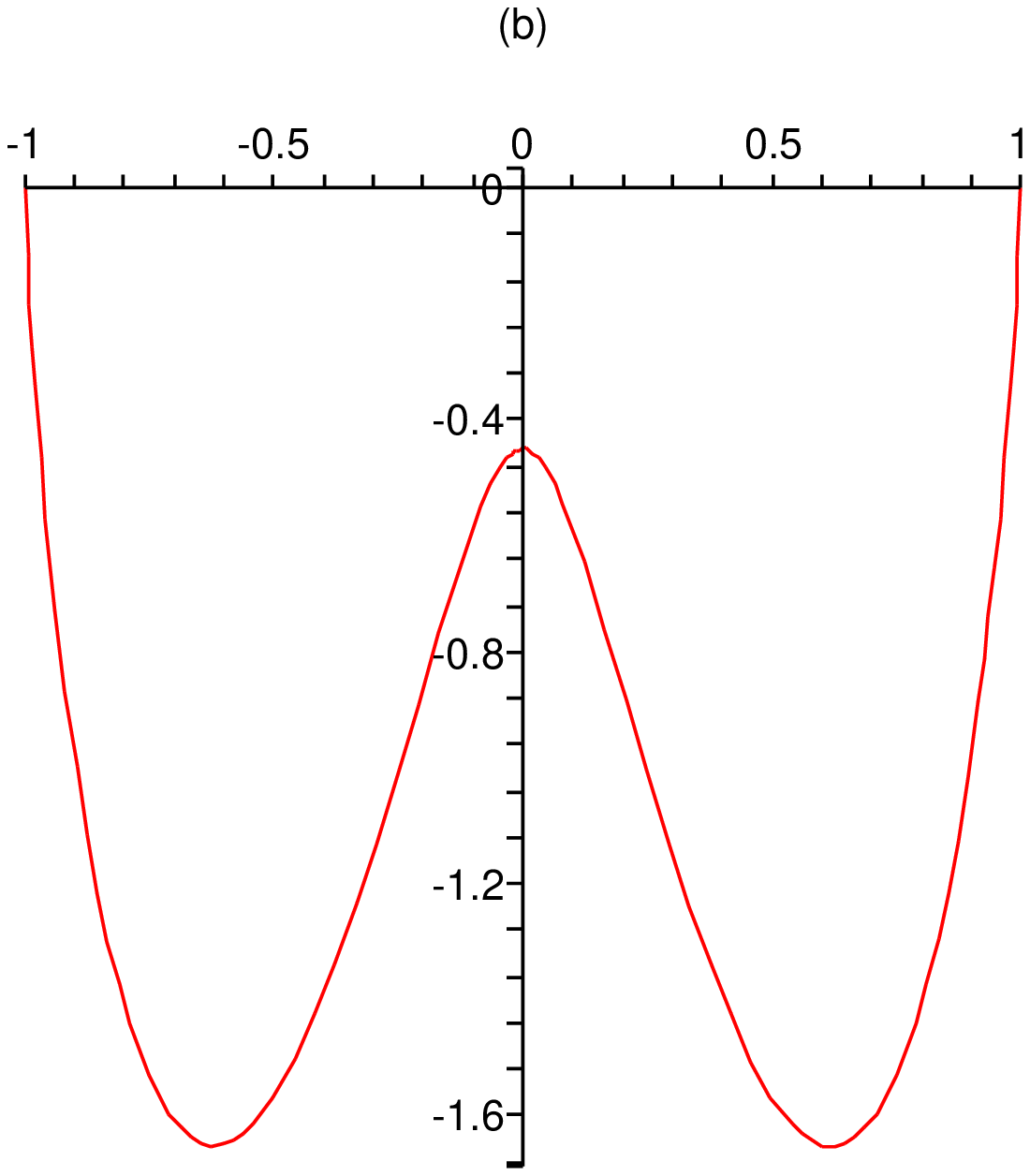}
    \end{center}
\begin{quote}
  \caption{\it The potential $-\cZ(\Phi,\ell)$ for
  a) $\ell^2=2\ell_c^2$ and for
  b) $\ell^2=30\ell_c^2$. The maximum and minimum of the
  potential is independent of $\ell$, however, the valley of the
  potential increases as $\ell$ increases.}
  \end{quote}
  \label{Fig3}
\end{figure}

We are interested in studying  the tunneling from the local
minimum $\Phi=0$ to the global minimum $\Phi=-1$ of the potential
$\cZ$. In terms of the new field \reef{newphi}, the action
\reef{SE} becomes \beqa S_E&=&\frac{T_{p-1}}{\sqrt{\alpha'}}\int
d^{p+1}x \sqrt{\cZ^2(\Phi,\ell)+ \alpha'
\prt_a\Phi\prt^a\Phi}\,\,\,.\nonumber\eeqa Following
\cite{Coleman}, one has to find a  solution with maximal
symmetry, \ie $\Phi=\Phi(r)$ where $r^2=x_ax^a$, and with the
boundary conditions $ \dot{\Phi}(0)=0$ and $
\Phi(\infty)=\dot{\Phi}(\infty)=0$. For such a solution the above
action becomes \beqa S_E&=&\frac{T_{p-1}}{\sqrt{\alpha'}}A_p
I_E\label{ES}\,\,,\eeqa where $A_p$ is the area of the $p$-sphere
with radius 1, \ie $A_p=2\pi^{(p+1)/2}/\Gamma(1/2+p/2)$,  and the
effective one-dimensional
action is \beqa I_E&=&\int_0^\infty dr\,\cL\nonumber\\
&=&\int_0^\infty dr\,
r^p\sqrt{\cZ^2+\alpha'\dot{\Phi}^2}\,\,.\eeqa  The Euler Lagrange
equation of motion is\beqa
\frac{\alpha'\ddot{\Phi}Z^2}{(Z^2+\alpha'\dot{\Phi}^2)^{3/2}}+
\frac{p}{r}\frac{\alpha'\dot{\Phi}}{\sqrt{Z^2+\alpha'\dot{\Phi}^2}}
&=&\frac{ZZ'}{\sqrt{Z^2+\alpha'\dot{\Phi}^2}}(1+\frac{\alpha'
\dot{\Phi}^2}{Z^2+\alpha'\dot{\Phi}^2}) \,\,.\eeqa The second term
on the left hand side  represents  friction term which decreases
as time progresses. It is easier to work in the Hamiltonian
formulation to study this nonlinear system. The equations of
motion in the Hamiltonian formulation are \beqa
\dot{\Pi}&=&\cZ'\sqrt{1-\Pi^2/\alpha'}-\frac{p}{r}\Pi\,\,,\nonumber\\
\dot{\Phi}&=&\frac{\cZ\Pi/\alpha'}{\sqrt{1-\Pi^2/\alpha'}}
\labell{eom}\,\,.\eeqa where $\Pi$ is related to the conjugate
momentum of the field $\Phi$ as\beqa \Pi&=&
\frac{\alpha'\dot{\Phi}}{\sqrt{\cZ^2+\alpha'\dot{\Phi}^2}}\,\,.\label{pi}\eeqa
Note that  $\Pi$ is bounded between $-\sqrt{\alpha'}$ and
$\sqrt{\alpha'}$. The bounce solution is a solution of the above
equations \reef{eom} with boundary conditions $\Pi(0)=0$ and
$\Phi(\infty)=\Pi(\infty)=0$.

Because of  the friction term  for $p>0$, system is not
conservative. It is still useful to consider the energy: \beqa E
\,\,\equiv\,\,\frac{1}{r^p}H&=&\frac{1}{r^p}\left(\dot{\Phi}
\frac{\prt\cL}{\prt\dot{\Phi}}-\cL\right)\nonumber\\&=&-\cZ\sqrt{1-\Pi^2/\alpha'}\,\,.\eeqa
Using \reef{eom}, one can easily show that the lost of energy is
\beqa \dot{E}&=&-\frac{p}{\alpha'
r}\frac{\cZ\Pi^2}{\sqrt{1-\Pi^2/\alpha'}}\,\,=\,\,-F\dot{\Phi}\,\,,\eeqa
where $F=p\Pi/r$ is the friction.

We now show that there is an "overshoot" point for any brane
separation. To see this, one has to study the behavior of the
system near the true vacuum.  Around the true vacuum, $\cZ\sim 0$,
the equations \reef{eom} are decoupled, \ie \beqa
\dot{\Pi}&=&\cZ'\sqrt{1-\Pi^2/\alpha'}-\frac{p}{r}\Pi\,\,,\nonumber\\
\dot{\Phi}&\sim&0\label{eom1}\,\,,\eeqa where $\cZ'$ is a very
large number near the true vacuum.

We first consider $p=0$.  In this case, the above  equations have
the following analytic solution: \beqa
\Pi(r)\,\,=\,\,\sqrt{\alpha'}\sin(\cZ'r/\sqrt{\alpha'})&;&\Phi(r)\sim-1\,\,,
\qquad\qquad  {\rm for}\,\,
r<\frac{\pi\sqrt{\alpha'}}{2\cZ'}\nonumber\\
\Pi(r)\,\,=\,\,\sqrt{\alpha'}&;&\Phi(r)\sim-1\,\,, \qquad\qquad
{\rm for}\,\,
r\geq\frac{\pi\sqrt{\alpha'}}{2\cZ'}\,\,.\label{solution}\eeqa At
the time $R/\sqrt{\alpha'}=\pi/(2\cZ')$, $\Pi$ reaches to its
maximum value and stays there  forever. In this case,
conservation of energy guarantees the existence of the bounce
solution, and the above result indicates that the initial value
$\Phi(0)=\Phi_0$ of the bounce solution is larger than $-1$. One
can find $\Phi_0$ by using the conservation of energy, \ie
$\cZ(\Phi_0)=\cZ(\Phi=0)$, or, in terms of old field, \beqa
U(\phi_0)\sqrt{1+\frac{\ell^2}{(2\pi)^2\alpha'}\phi_0^2}&=&U(0)\,\,.\eeqa
The coefficient $B$ for the bounce solution is \beqa
B&=&\frac{T_{-1}}{\sqrt{\alpha'}}\left(\int_0^\infty dr
\sqrt{\cZ^2+\alpha' \dot{\Phi}^2}-\int_0^\infty dr  U(0)\right)\nonumber\\
&=&T_{-1}\int_{\phi_0}^0
\frac{\sqrt{U^2(\phi)\left(1+\frac{\ell^2}{(2\pi)^2\alpha'}\phi^2\right)-U^2(0)}}
{\sqrt{1+\frac{\ell^2}{(2\pi)^2\alpha'}\phi^2}}d\phi\,\,,\label{B0}\eeqa
where we have used conservation of energy to write $\dot{\Phi}$
in terms of energy and tachyon potential. Note that, for large
$\ell$, the integral is independent of the brane separation, \ie
$B\sim T_{-1}\int_{-1}^0d\Phi=T_{-1}=\ell_cT_0$. We will show in
the next section that this property  is hold even for $p>0$ cases.

For $p>0$ cases, the qualitative form of the solution of equations
\reef{eom1} is the same as \reef{solution}. However, because of
the  friction term $p\Pi/r$ in \reef{eom1}, $\Pi$  reaches  to its
maximum value in a bit longer than $R/\sqrt{\alpha'}=\pi/(2\cZ')$.
Then particle stays there for arbitrary long period of time by
fine tuning the initial value of $\Phi$, \ie the closer $\Phi$ to
$-1$, the longer  particle stays around $\Phi=-1$. After staying
for  arbitrary  period of time at $\Phi\sim -1$, particle moves
off this point with negligible friction toward the false vacuum.
Therefore, there is an "overshoot" point for any brane
separation. In the next subsection we use the thin-wall
approximation \cite{Coleman} to find the bounce solution for
large brane separation.

\subsection{The thin-wall approximation}

We now find the bounce solution and calculate the coefficient B
for large brane separation, and for $p>0$. The qualitative form
of the bounce solution for large $\ell$ is the following: In
order not to lose too much energy due to the friction, we must
choose the initial position of the particle very close to $-1$.
The particle then stays close to $-1$ until some very large time,
$r=R$. Near time $R$, the particle moves quickly through the
valley of the potential $-\cZ$, and slowly comes to rest at
$\Phi=0$ at time infinity, \ie \beqa \Phi&=&-1\,,\qquad\qquad
r<<R\nonumber\\
&=&\Phi_1(r-R)\,,\,\,\,\,\,\, r\simeq R \label{approx}\\
&=&0\,,\qquad\qquad\,\,\,\,\,\, r>>R\nonumber\eeqa where
$\Phi_1(r-R)$ is the solution that starts at $\Phi=-1$ with
negligible friction term and ends at $\Phi=0$. Since the friction
is neglected for this part, we can use conservation of energy to
find this solution. Using the facts that energy is
$E=-\cZ\sqrt{1-\Pi^2/\alpha'}$ and $\cZ(-1)=0$, one observes that
$\Pi=\sqrt{\alpha'}$ during this period. From equation \reef{pi},
one realizes that velocity of the particle should be infinite,
\ie $ \dot{\Phi}_1=\delta(r-R)$. The only thing missing from this
description is the value of $R$ which can be obtained by
variational computation:\beqa
B&=&\frac{T_{p-1}}{\sqrt{\alpha'}}A_p\left(\int_0^\infty dr
r^p\sqrt{\cZ^2+\alpha' \dot{\Phi}^2}-\int_0^\infty dr r^p \cZ(0)\right)\nonumber\\
&=&\frac{T_{p-1}}{\sqrt{\alpha'}}A_p\left(\int_0^R dr
r^p(\cZ(-1)-\cZ(0))+\alpha'R^p\int_{-1}^0 \frac{d\Phi}{\Pi}\right)\\
&=&\frac{T_{p-1}}{\sqrt{\alpha'}}A_p\left(-\frac{\epsilon}{p+1}R^{p+1}+
\sqrt{\alpha'}R^p\right)\nonumber\eeqa where
$\epsilon=\cZ(0)-\cZ(-1)=\sqrt{2}/\pi$. Varying with respect to R,
one obtains \beqa \frac{dB}{dR}=0& \Longrightarrow &
R_0=\frac{p}{\epsilon}\sqrt{\alpha'}\,=\,\frac{p}{2}\ell_c\,\,.\label{R}\eeqa
Thus the radius of the nucleated bubble is independent of the
brane separation for large $\ell$. This is unlike the result in
\cite{koji} that uses two-derivative truncation of the tachyon
action. We note that the energy lost for the solution
\reef{approx} is exactly equal to the difference energy between
true and false vacuum, \ie
 \beqa
\Delta E&=&\int_0^{\infty}dr \,F\dot{\Phi}
\,=\,\frac{p\sqrt{\alpha'}}{R_0}\,=\,\epsilon \eeqa where we have
replaced the radius of nucleated bubble from \reef{R}.

The condition for validity of the thin-wall approximation is
\cite{Coleman}\beqa \mu
\left(\frac{R_0}{\sqrt{\alpha'}}\right)>>1\label{condition}\eeqa
where $\mu$ is the scale parameter of the theory. To find this
parameter, we need to study the behaviour of the potential around
its maximum, \ie\beqa
\cZ-\cZ_{top}&=&\frac{1}{2}\mu(\delta\Phi)^2+\cdots\eeqa The
dimensionless potential around the top of the hill is\beqa
\cZ-\cZ_{top}&\sim&\frac{1}{2}\frac{\prt^2\cZ}{\prt\phi^2}(\delta\phi)^2\nonumber\\
&=&\frac{1}{2U^2}\frac{\prt^2\cZ}{\prt\phi^2}(\delta\Phi)^2\eeqa
For tachyon potential $U=(\sqrt{2}/\pi)e^{-\phi^2/4}$, and for
large brane separation, \ie $\ell>>\sqrt{\alpha'}$, the maximum
of the $D\bar{D}$ potential $\cZ$ is at $\phi\sim\sqrt{2}$. For
this case one finds \beqa
\mu&=&\frac{1}{U^2}\frac{\prt^2\cZ}{\prt\phi^2}|_{\phi=\sqrt{2}}\,\sim\,
\left(\frac{\sqrt{e}}{2}\right)\frac{\ell}{\sqrt{\alpha'}} \eeqa
For other tachyon potential,  the numerical factor above changes.
Therefore,\beqa
\mu\left(\frac{R_0}{\sqrt{\alpha'}}\right)&\sim&\frac{p\pi}{2}
\sqrt{\frac{e}{2}}\frac{\ell}{\sqrt{\alpha'}}\,\,,\eeqa for large
brane separation, the condition \reef{condition} is satisfied
which means our thin-wall approximated solution \reef{approx} is
valid. Note that, for the minimum value of the branes separation
where there are false and true vacuums, \ie $\ell=\ell_c$, the
right hand side is $p\pi^2\sqrt{e}/2\sim 8p$ which is larger than
one.

Finally, the decay width in the thin-wall approximation
becomes\beqa
B&=&2T_pA_p\left(-\frac{1}{p+1}R_0^{p+1}+\frac{1}{p}R_0^{p+1}\right)\nonumber\\
&=&\frac{2T_{p}A_p}{p(p+1)}R_0^{p+1}\eeqa which is independent of
brane separation $\ell$! Note that, the above result for $p=0$
gives $B=T_{0}\ell_0$ which is consistent with the result in
\reef{B0} for large brane separation. This unusual result may
indicate that the assumption that the transverse scalar fields of
branes are fixed, \ie no throat formation,   is not a valid
assumption. Or, it may indicates that  the form of the effective
action is not given  by the tachyon DBI action when branes
separation is larger than $\ell_c$.

To find  the geometrical picture of the bubble, we note that in
the  approximated solution \reef{approx}, inside the Euclidean
sphere is filled with the true vacuum where the tachyon is
completely condensed and so the brane-anti-brane are disappeared.
Outside the sphere, on the other hand,  is filled with the false
vacuum where the tachyon does not condensed and so the
brane-anti-brane do exist. The geometrical picture of
brane-anti-brane after decay is then the original brane-anti-brane
in which a spherical hole with radius \reef{R} is created at the
center of each brane, see fig.4. The radius of the bubble is
independent of branes separation! These two holes may connect to
each other by forming a throat \cite{callen,koji}. To study this
throat formation, one has to release the assumption that the
transverse scalars of the brane are fixed, and find a solution
which includes both tachyon and the transverse scalar fields. In
other words, one has to find  a solution of the coupled equation
of motion of massless scalar and tachyon which represents both
tachyon bounce and the throat formation. One expects in this case
that the radius of the hole depends on the branes separation, the
larger the branes separation, the larger the radius of the hole.
We leave this study for the future.

\begin{figure}
  \begin{center}
 \includegraphics{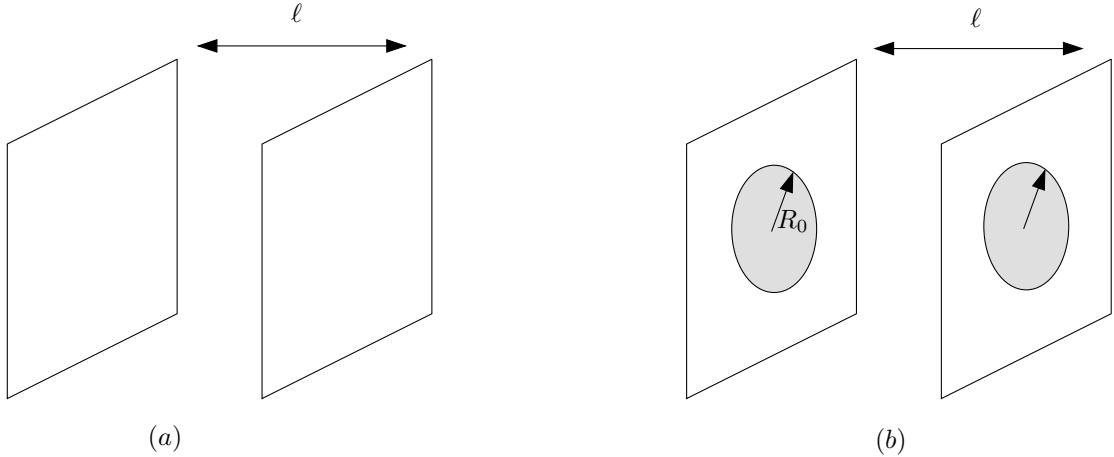}
    \end{center}
\begin{quote}
  \caption{\it The time evolution of $D_2$-$\bar{D}_2$ system for condensation of tachyon into bounce solution.
  (a) Before decay.
  (b) After decay.}
  \end{quote}
  \label{Fig4}
\end{figure}

 {\bf Acknowledgement}: We would like to thank M.S. Costa, A.
 Ghodsi, K. Hashimoto, K. Javidan, G.R. Maktabdaran  and J. Penedones for
 discussions. M.R.G would like to thank Perimeter Institute
for hospitality.


\end{document}